\documentclass[pre,twocolumn,amsmath,amssymb,superscriptaddress]{revtex4}
\usepackage{graphicx}
\usepackage{bm}
\usepackage{amsmath}
\usepackage{amsfonts}
\usepackage{amssymb}%

\newcommand{\be}{
\begin{equation}
}
\newcommand{\ee}{l
\end{equation}
}
\newcommand{\beq}{
\begin{eqnarray}
}
\newcommand{\eeq}{
\end{eqnarray}
}
\begin{document}
\title{ Inverse square L\'evy walks are not  optimal search strategies for $d\ge 2$\\
 }

The central result of our letter \cite{Levernier:2020aa} is that (i) the capture rate $\eta$ of Levy walks with Poisson distributed targets goes linearly with the target density $\rho$ for all values of the Levy exponent $\alpha$ in space dimension $d\ge 2$. This  contradicts   results in \cite{Viswanathan:1999a},  and has  important consequences: (ii) the optimal gain $\eta_{\rm max}/\eta$ achieved by varying $\alpha$ is bounded  in the limit $\rho\to 0$ so that tuning $\alpha$ yields a marginal gain ; (iii) the optimum is realized for a range of $\alpha$ and is controlled by the model-dependent parameters $a$ (detection radius),  $l_c$ (restarting distance) and $s$ (scale parameter) (Fig.1). 

First, and most importantly,  \cite{buldy2000} states that our main result  (i) is correct, thereby acknowledging that the determination of $\eta$  in \cite{Viswanathan:1999a} is wrong.

Second, \cite{buldy2000} opposes that  claim (iii) is not new, because earlier publications reported that optimal Levy strategies can be realized for $\alpha\not=1$. We did acknowledge such {\it observations}  in \cite{Levernier:2020aa}, where we in fact   
  {\it show} that they result from the linear scaling of $\eta$ with $\rho$ for  $d\ge 2$ ; this is novel.
\begin{figure}[h!]
\includegraphics[width=7.5cm]{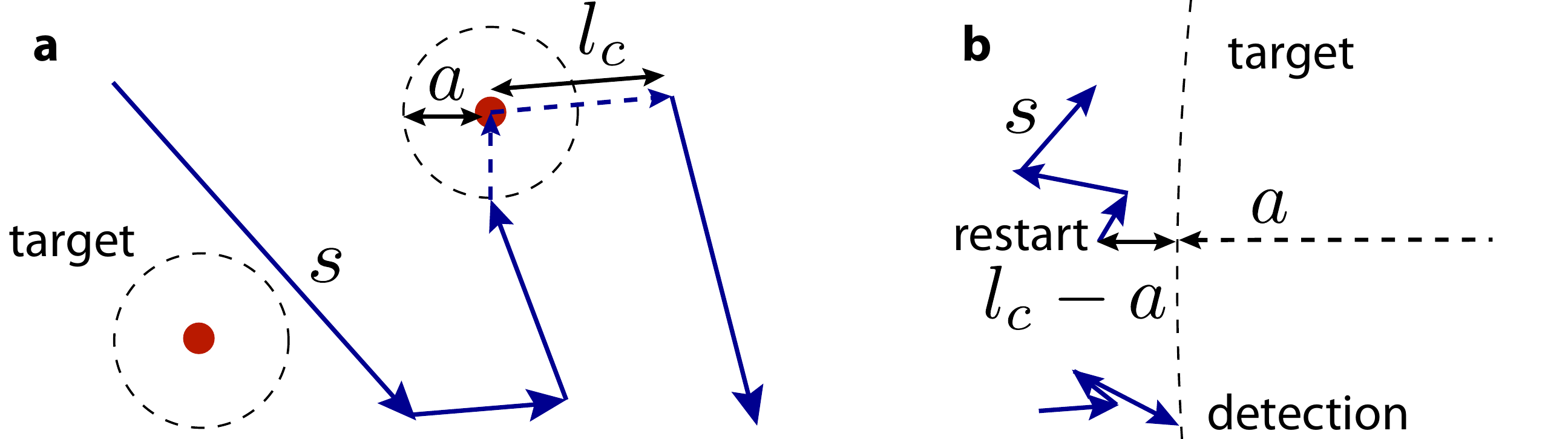}
\caption{ The L\'evy walk search model in the generic $2d$ case ({\bf a}). Inverse square Levy walks are optimal only in the singular $1d$ limit $l_c-a\ll a$ and  $s\ll a$ ({\bf b}).}
\label{fig1}
\end{figure}

Last,   \cite{buldy2000} disputes  claim (ii). Technically,  claim (ii) is  correct and by no means compromised by \cite{buldy2000}. It states that for {\it fixed }values of $s,l_c$, the optimal gain $\eta_{\rm max}/\eta$  is bounded  when $\rho\to 0$. This comes from the linear scaling of $\eta$ with $\rho$ (Eq. (5) in \cite{Levernier:2020aa}, whose validity is  acknowledged  by \cite{buldy2000}), and is  {\it independent} of any determination of $K_d(\alpha,s,l_c)$.  In \cite{Levernier:2020aa},  Eq.(3)  is used only to derive the scaling of $\eta$ with $\rho$ ; we make no prediction regarding $K_d(\alpha,s,l_c)$. Attempting to deduce $K_d(\alpha,s,l_c)$ from Eq.(3) is   the initiative of \cite{buldy2000}, not ours ; in fact, we  agree that Eq.(3) is unsuitable to study $l_c\to a$, which falls out of the validity regime   given in \cite{Levernier:2018qf}. This is certainly not a problem in \cite{Levernier:2020aa} as argued by  \cite{buldy2000}, simply because we nowhere aimed at determining  $K_d(\alpha,s,l_c)$.

Finally, the only aspect in (ii) that  \cite{buldy2000}   disputes is  rethoric -- our qualification of the optimum as marginal. The comment is based only  on the analysis  of  the singular limit $s\to0$ and $l_c\to a$, which can indeed lead to arbitrary large values of  $\eta_{\rm max}/\eta$ for $\alpha\to 1$.  This  is actually a mere $1d$ limit  (Fig. 1), as noted in \cite{Levernier:2020aa} ; it is thus expected, and  consistent with our findings, to recover the $1d$ optimum.This  by no means contradicts  claim (ii) of boundedness when $\rho\to 0$ {\it for fixed} $s,l_c$.     Last, we summarize the conditions of optimality (CO) of inverse square Levy walks for  $d\ge2$ : \\
-	upon each capture event, a spherical target reappears infinitely fast at the  same position\\
-	the searcher starts the new search infinitely close to the target boundary ($l_c-a\ll a$).\\
-	the typical scale of its displacements is infinitely smaller than the target ($s\ll a$).\\
If {\it any} of these conditions is not met, $\alpha= 1$ is not optimal. Given that  $s$ and $l_c$ are system-dependent parameters with arbitrary values, CO are generically not met and our conclusion that inverse square Levy walks are not  optimal is justified. Additionally,  if $l_c,s$ are allowed to vary, as done in  \cite{buldy2000}, the obvious optimal strategy is $l_c = a$, leading to immediate recapture of the same target ; the limit $l_c\to a^+$ in CO  is thus artificial.

 To our knowledge, CO have never been stated explicitly, nor verified in any experimental system.  Given that CO are a mere $1d$ limit of the  problem, the claim that  \cite{buldy2000}  restores  the optimality of $\alpha= 1$ for  $d\ge2$ is  unfounded ;  given that  \cite{buldy2000} acknowledges that the scaling of $\eta$ with $\rho$ is wrong in \cite{Viswanathan:1999a}, stating that  \cite{buldy2000} restores the validity of \cite{Viswanathan:1999a} is also unfounded.




\vspace{0.5cm}
\noindent Nicolas Levernier, Olivier B\'enichou , Rapha\"el Voituriez \\
Laboratoire de Physique Th\'eorique de la Mati\`ere Condens\'ee, CNRS /UPMC, 4 Place Jussieu, 
Paris, France

\vspace{0.5cm}

\noindent  Johannes Textor\\
Institute for Computing and Information Sciences
Radboud University
Nijmegen, The Netherlands

\end{document}